\documentclass[twocolumn,aps,prd,nofootinbib,superscriptaddress,showkeys]{revtex4-1}
\usepackage[colorlinks=false,linkcolor=black,citecolor=black]{hyperref}
\usepackage{graphicx}
\usepackage{subfig}
\usepackage{amstext,amsmath,amsfonts,amssymb,amsthm}
\usepackage[format=plain,justification=centerlast]{caption}
\usepackage{hyperref} 
\usepackage{color}

\usepackage{verbatim}
\def\app#1{{Appendix~\ref{#1}}}
\def\sec#1{{Section~\ref{#1}}}
\def\eq#1{{Eq.~(\ref{#1})}}

\def\fig#1{{Fig.~\ref{#1}}}

\def\dd{\mathrm{d}}
\def\be{\begin{equation}}
\def\ee{\end{equation}}
\def\bes{\begin{eqnarray}}
\def\ees{\end{eqnarray}}
\def\ba{\begin{align}}
\def\ea{\end{align}}
\def\bwt{\begin{widetext}}
\def\ewt{\end{widetext}}

\def\pp{\partial}
\def\nn{\nonumber}

\begin{document}

\title{Thermodynamics of Gravity in Local Frames}
\date{\today}
\author{Suprit Singh}
\email{ssingh2@physics.du.ac.in}
\affiliation{Department of Physics, Indian Institute of Technology Delhi, Hauz Khas, New Delhi, India 110016.}

\begin{abstract}
We probe the thermodynamic structure of gravity at local scales. In any general curved spacetime, it is possible to transform to a local inertial frame at any point such that the metric is flat up to quadratic order where the curvature at that point comes in when the metric is written in Riemann normal coordinates. We consider local Rindler observers in that patch and hence the local Rindler horizon. In doing so, we find that the local horizons are also hot provided the $aL>>1$ which can always be satisfied. 
\end{abstract}


\maketitle

\section{Introduction} \label{sec:intro}

We can probe the nature of quantum spacetime by following a `top-down' approach, that is, by learning key lessons from the way classical gravity interacts with quantum fields leading to thermodynamics of horizons (for more insights see \cite{lessonsTP}). These studies suggest that \emph{gravity is an emergent phenomenon} arising out of the description of spacetime at large length scales compared with some critical length scale which, possibly - but not necessarily - could be the Planck length. This is similar to the way one thinks of fluid dynamics as an emergent phenomenon with its own dynamical variables and corresponding equations of motion that are valid at scales much bigger than the typical intermolecular separations.\\
\\
There is one key link between the microscopic and the macroscopic aspects that can be exploited. This is the \emph{thermal phenomenon} involving the concept of temperature and transfer of heat energy and is an evidence ($\grave{\mathrm{a}}$ la Boltzmann) of an underlying microstructure which is well known for fluids. We have a similar connection in gravity via the \emph{horizons}. Horizons are inevitable structures in any theory of gravity obeying the principle of equivalence and we can always have nontrivial null surfaces which block the information from a certain class of observers. Thus, it is \textit{always an observer dependent concept} and precisely this idea of local horizons and their relation with thermodynamics forms one of the basis for emergent gravity paradigm.\\
\\
A well known and simplest case of spacetime with a horizon is of an accelerated or Rindler observer (the results also hold for cases in which the near-horizon geometry can be approximated as Rindler geometry) which exhibits a thermodynamic character. The temperature of the horizon is an observable effect in a way that the vacuum state of the quantum fields in a Minkowski spacetime appears thermal to an accelerated Rindler observer with a temperature proportional to the magnitude of the acceleration. This is the well-documented Unruh effect \cite{crispino,tpreview}.The notion and the expressions for entropy and temperature as seen by the Rindler observer are well defined. But this is established in the case of spacetimes with global horizons. The question arises whether one define a temperature and look at thermal aspects through quantum fields in a local region around any arbitrary spacetime event \cite{tpreview}. We show here that it can indeed be done.\\
\\
The plan of the paper is as follows. In~\sec{standard}, we briefly review the quantum field theory for a scalar field as seen from an accelerated frame and calculate the Bogoliubov coefficients (numbers relating the inertial and Rindler field modes) on the null surface getting the standard thermal aspects. We (in~\sec{goinglocal}) introduce the notion of local Rindler observers in the local inertial frame which can always be set up at any event on a generic spacetime manifold. Once caged in the local patch, the Bogoliubov coefficients are again evaluated but in the restricted domain in~\sec{caged} and we see that thermal aspects are indeed recovered in a sensible limit. 

\section{A look at the scalar field from accelerated perspective}
\label{standard}
In a standard Rindler (accelerated observer in flat spacetime) setting, we shall work with the metric of the form
\begin{equation*}
ds^2 = a^2\rho^2 d\tau^2 - d\rho^2 - dx^2 - dy^2
\end{equation*}
which can immediately be transformed to
\begin{equation}
ds^2 = e^{2a\xi}(d\tau^2 - d\xi^2) - d\bf{x}_\bot^2
\end{equation}
with the following definition:
\begin{equation*}
\xi \equiv a^{-1} \ln a\rho.
\end{equation*}
The Klein-Gordon equation for the scalar field in the above metric then reads,
\begin{equation*}
\left[\pp_\tau^2 -\pp_\xi^2 - e^{2 a \xi}\pp_\bot^2  + m^2 e^{2 a\xi}\right]\Phi(t,\xi, \bf{x}_\bot) = 0 
\end{equation*}
Using the symmetry arguments, we can take the positive frequency solution to be
\begin{equation*}
\Phi^+ \sim e^{-i\omega\tau+i \bf{k}_\bot\cdot\bf{x}_\bot}f(\xi)
\end{equation*}
which then gives the equation for $f(\xi)$ to be
\begin{equation}
\left[-\frac{d^2}{d\xi^2}+e^{2a\xi}(\bf{k}^2_\bot+m^2)\right]f(\xi) = \omega^2 f(\xi).
\end{equation}
We shall now break away from the direct solution of this equation which are essentially modified Bessel functions of the second kind~\cite{crispino}. Instead, let us take the limit $\xi\rightarrow-\infty$, that is, we will look at this equation and its solution near the future null surface. This can be done, since we need to compute the Bogoliubov coefficients connecting the field modes in two different frames, and the inner product which defines them is independent of the choice of the hypersurface on which it is defined. The choice of working near the null surface makes the computations simpler as then the transverse modes decouple and it is as good as working in a 1+1 spacetime. The scalar field in the right riddler wedge then can be expressed as
\begin{equation}
\Phi_+(v) = \int_0^\infty \dd\omega \left[\hat{a}_\omega g_\omega(v) + \hat{a}^\dagger_\omega g^*_\omega(v) \right]\:\:\:(\rm{for}\:V > 0)
\end{equation}
taking only the left moving modes at the moment with $v = t +\xi$ and with the positive frequency solution:
\begin{equation}
\label{gomegav}
g_\omega(v) = \frac{1}{\sqrt{4\pi\omega}} e^{-i\omega v}.
\end{equation} 
Also,
\begin{equation}
\Phi(t,z) = \Phi_-(U) + \Phi_+(V); \:\:\: U = t - z,\,\,V=t+z.
\end{equation}
and 
\begin{equation}
\Phi_+(V) = \int_0^\infty \dd k\left[\hat{b}_k f_k(V) + \hat{b}^\dagger_k f^*_k(V)\right]
\end{equation}
with $f_k(V) = (4\pi k)^{-1/2}e^{-ikV}$. Expanding $g_\omega(v)$ in terms of the Minkowski modes, we have,
\begin{equation}
\Theta(V) g_\omega(v) = \int_0^\infty \frac{\dd k}{\sqrt{4\pi k}} (\alpha_{\omega k} e^{-ikV}+ \beta_{\omega k}e^{ikV}). 
\end{equation}
Then,
\begin{equation}
\label{alpha}
\alpha_{\omega k} = \sqrt{4\pi k}\int_0^\infty \frac{\dd V}{2\pi}g_\omega(V)e^{ikV}
\end{equation}

\begin{equation}
\label{beta}
\beta_{\omega k} = \sqrt{4\pi k}\int_0^\infty \frac{\dd V}{2\pi}g_\omega(V)e^{-ikV}
\end{equation}
Their computation is straightforward using \eq{gomegav} and substituting for $V = a^{-1}e^{av}$, 
\begin{align}
\alpha_{\omega k} &= \sqrt{\frac{k}{\omega}} \int_{-\infty}^{\infty}\frac{dv}{2\pi}\,e^{-i\omega v} e^{av}e^{i\frac{k}{a}e^{av}}\nonumber\\
&= \sqrt{\frac{k}{\omega}} \left(-\frac{ia}{k}\right)\lim_{\mu \to 1}\pp_\mu \int_{-\infty}^{\infty}\frac{dv}{2\pi}\,e^{-i\omega v}\, e^{i\frac{k}{a}\mu e^{av}}\nonumber\\
&= \frac{e^{\pi\omega/a}}{2\pi a}\sqrt{\frac{\omega}{k}}\left(\frac{k}{a}\right)^{i\omega/a}\Gamma\left(-\frac{i\omega}{a}\right)
\end{align}
\\
Similarly we can also evaluate $\beta_{\omega k}$ as,
\begin{equation}
\beta_{\omega k} = -\frac{e^{-\pi\omega/a}}{2\pi a}\sqrt{\frac{\omega}{k}}\left(\frac{k}{a}\right)^{i\omega/a}\Gamma\left(-\frac{i\omega}{a}\right)
\end{equation}

\section{Going Local}
\label{goinglocal}

So much so for the global scenario. In the emergent gravity perspective as discussed earlier, we need to look at the local Rindler observers, which can be introduced at any event in space-time and use the horizon perceived by these Rindler observers to ascertain the local thermodynamical effects. To do this, we shall first begin by defining the notion of local Rindler observers and their coordinate system.\\
\\
In any generic spacetime, we can always introduce a local inertial frame (LIF) at any event say, $\mathcal{P}$ in terms of the Riemann normal coordinates (see ~\cite{rnc}), $X^a = (T,\bf{X})$ such that the point $\mathcal{P}$ lies at the origin of the LIF.  This frame is `locally inertial' in the sense that the deviation of the Minkowski metric is at quadratic order in coordinates and is determined by the curvature tensor. 
\begin{equation}
ds^2 = (1-\frac{1}{3}\mathcal{R}X^2)dT^2 - (1+\frac{1}{3}\mathcal{R}T^2)dX^2 +\frac{2}{3}\mathcal{R}TX\,dX\,dT
\end{equation}
\begin{figure*}[t!]
\includegraphics[scale=0.35]{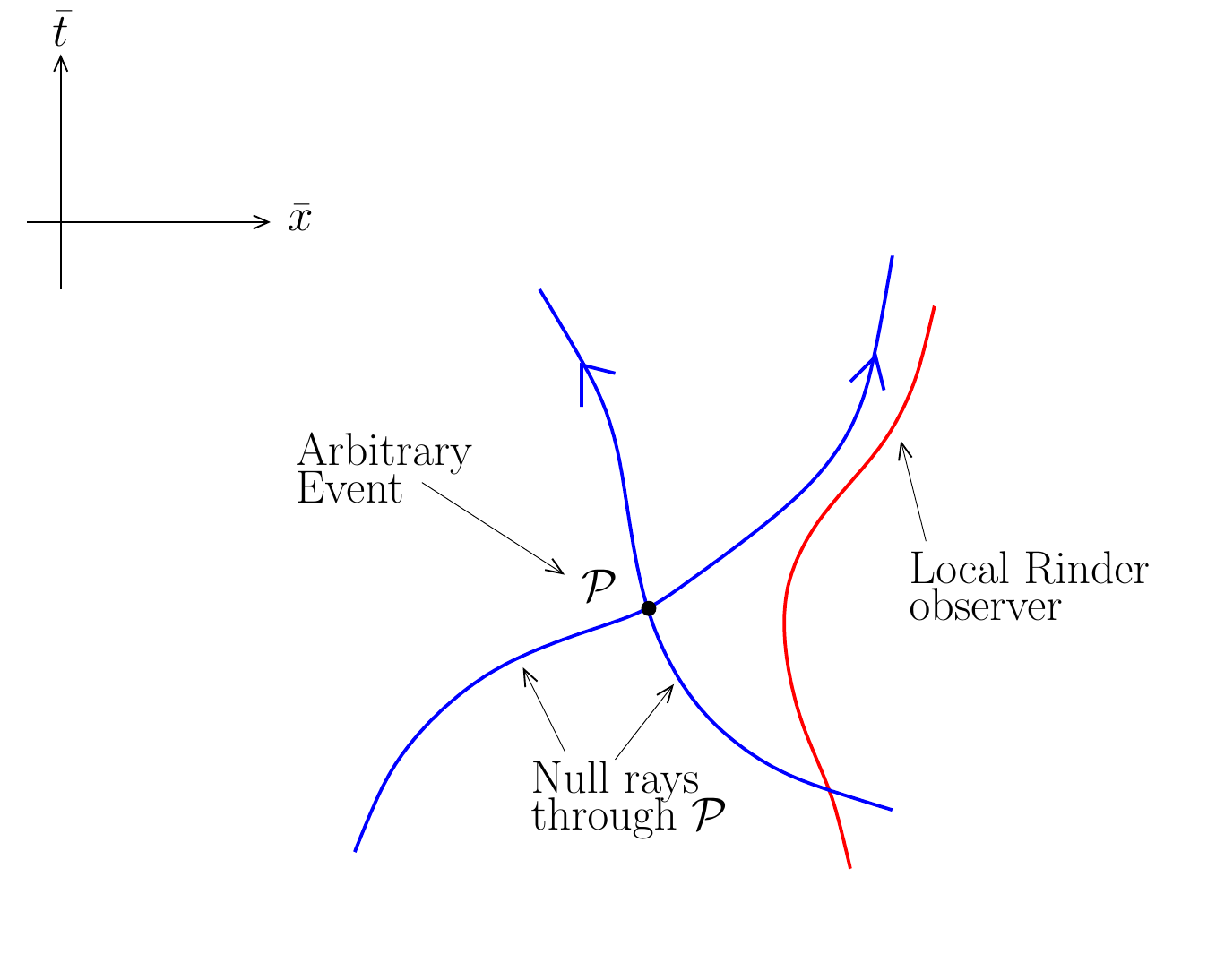}\hfill
\includegraphics[scale=0.35]{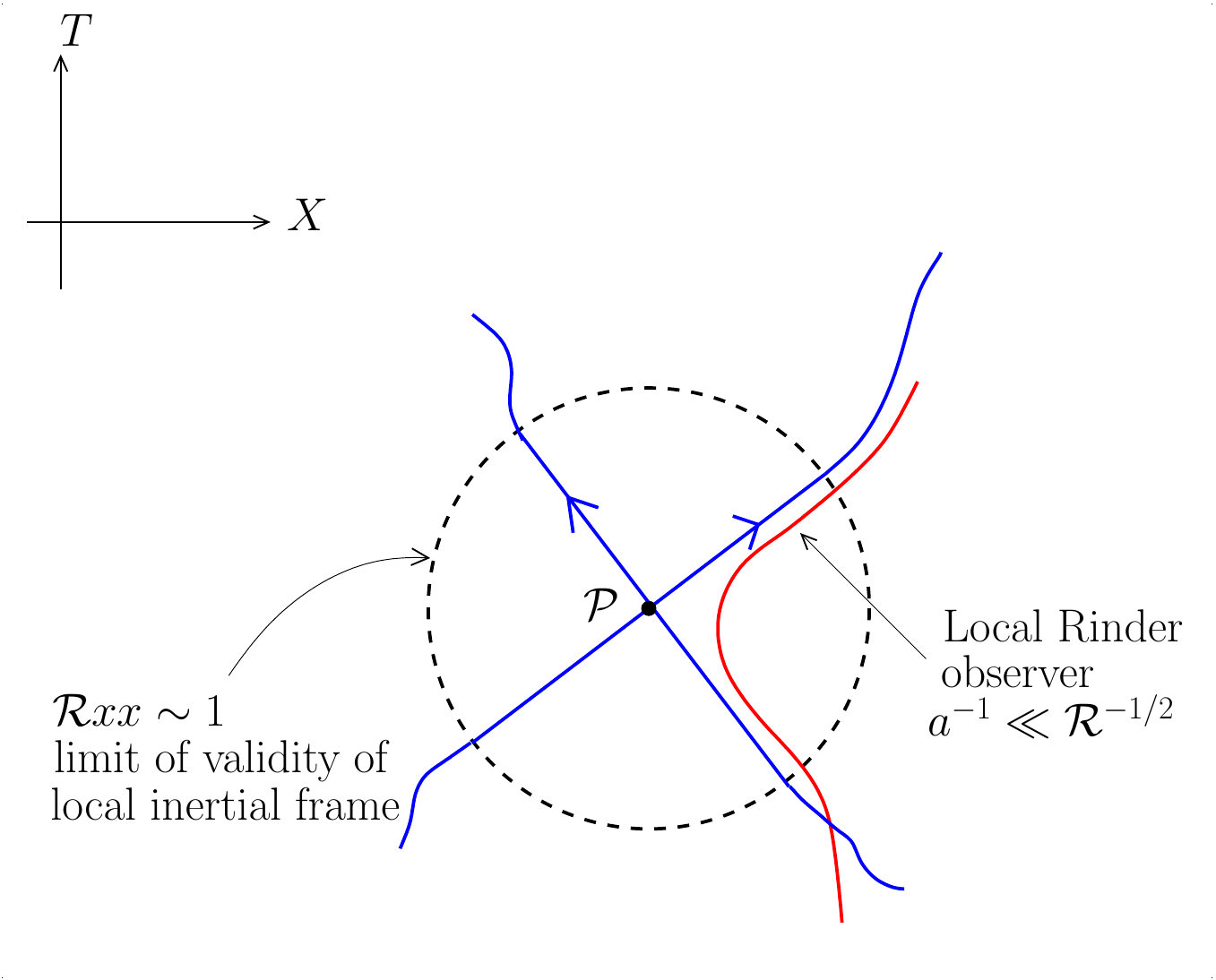}
\caption{Schematic illustration (adapted from T. Padmanabhan, Rep. Prog. Phys. 73 (2010) 046901)}{{\bf Left:} null rays around any event $\mathcal{P}$ in the generic spacetime; {\bf Right:} LIF and LRF at $\mathcal{P}$ with the null rays at 45 degrees and the trajectory of the local Rindler observer is a hyperbola very close to $T=\pm X$ lines acting as a local horizon to the Rindler observer.}
\label{liflrf}
\end{figure*}
We next transform from the LIF to a local Rindler frame (LRF) coordinates (see~\fig{liflrf}) by accelerating along the say, X-axis (of LIF) with an acceleration $a$ by the usual transformation. 
\begin{equation}
ds^2 = e^{2a\xi}\left[\left(1+\frac{\mathcal{R}}{3}\frac{e^{2a\xi}}{a^2}\right)d\tau^2 - d\xi^2\right]
\end{equation}
This LRF and its local horizon $\mathcal{H}$ will exist within a region of size $L\ll\mathcal{R}^{-1/2}$ (where $\mathcal{R}$ is a typical component of curvature tensor of the background spacetime) as long as $a^{-1}\ll\mathcal{R}^{-1/2}$. This condition can always be satisfied by taking the product $aL$ to be sufficiently large. With this setup, we can study quantum field theory in the local frames and relate the field vacuum states in the LRF and the LIF to ascertain the thermodynamic aspects. 

\section{Caging the observer}
\label{caged}
Once we have caged the observer inside a region of size $L$, it just requires to play the usual game and calculate the Bogoliubov coefficients (\eq{alpha}~and~\eq{beta}) but in the restricted domain so as to get the local counterparts.

\begin{equation}
\alpha_{\omega k}^{loc} = \sqrt{4\pi k}\int_0^{\bar{L}} \frac{\dd V}{2\pi}g_\omega(v)e^{ikV}
\end{equation}
\begin{equation}
\beta_{\omega k}^{loc} = \sqrt{4\pi k}\int_0^{\bar{L}} \frac{\dd V}{2\pi}g_\omega(v)e^{-ikV}
\end{equation}
\\
Looking first at the $\beta_{\omega k}^{loc}$ coefficient, we have

\begin{align*}
\beta_{\omega k}^{loc} &= \sqrt{4\pi k}\int_0^{\bar{L}}\frac{\dd V}{2\pi}\,g_\omega(v)e^{-ikV}\nonumber\\
&= \frac{1}{2\pi}\sqrt{\frac{k}{\omega}}\int_{-\infty}^L \dd v e^{av}e^{-i\omega v}e^{-i (k/a)\, e^{av}}\nonumber\\
&= \frac{i a}{2\pi\sqrt{\omega k}}\lim_{\mu \to 1}\pp_\mu \int_{-\infty}^L \dd v\, e^{-i\omega v}e^{-i (k\mu/a)\, e^{a v}}
\end{align*}
\\
where we have made use of the coordinate transformation, $V = a^{-1}e^{av}$. After some straightforward algebra (see~\app{calcofbeta} for details), we get,

\begin{align}
\label{localbeta}
\beta_{\omega k}^{loc} = & -\frac{e^{-\pi\omega/2a}}{2\pi a}\sqrt{\frac{\omega}{k}}\left(\frac{k}{a}\right)^{\frac{i\omega}{a}}\gamma\left(-\frac{i\omega}{a},\frac{k}{a}e^{aL}\right)\nn\\
& - \frac{i}{2\pi a}\sqrt{\frac{\omega}{ k}}e^{-i\omega L}I_c + \frac{i}{2\pi\sqrt{\omega k}}e^{-i\omega L}e^{-ik/a\,e^{aL}}
\end{align}

As discussed in the previous section, it is physical to look at the large $aL$ limit and also to get corrections to the standard Rindler case. Under this limit,

\begin{align}
\beta_{\omega k}^{loc} \approx &-\frac{e^{-\pi\omega/2a}}{2\pi a}\sqrt{\frac{\omega}{k}}\left(\frac{k}{a}\right)^{\frac{i\omega}{a}}\Gamma\left(-\frac{i\omega}{a}\right) + \nn\\
&\frac{e^{-\pi\omega/2a}}{2\pi k}\sqrt{\frac{\omega}{k}} e^{-i\omega L}e^{-aL}e^{-k/a\,e^{aL}}\nonumber\\
&-\frac{i}{2\pi k}\sqrt{\frac{\omega}{ k}}e^{-i\omega L} e^{-aL}e^{-i\,k/a\,e^{aL}} + \frac{i}{2\pi\sqrt{\omega k}}e^{-i\omega L}e^{-ik/a\,e^{aL}}\nonumber\\
=&-\frac{e^{-\pi\omega/2a}}{2\pi a}\sqrt{\frac{\omega}{k}}\left(\frac{k}{a}\right)^{\frac{i\omega}{a}}\Gamma\left(-\frac{i\omega}{a}\right) \nn\\
&-\frac{i}{2\pi k}\sqrt{\frac{\omega}{ k}}e^{-i\omega L} e^{-aL}e^{-i\,k/a\,e^{aL}} 
\end{align} 
\\
where we the second term is eliminated in comparison with the third term. The last term is spurious in that it oscillates rapidly for large $aL$ and it is eliminated in the standard Rindler case as well by adding a small imaginary part to the $k$ so as to make it convergent and vanish on the average. On similar lines, we can also calculate $\alpha_{\omega k}$ in the large $aL$ limit which turns out to be,
\begin{align}
\alpha_{\omega k}^{loc} = &\frac{e^{\pi\omega/2a}}{2\pi a}\sqrt{\frac{\omega}{k}}\left(\frac{k}{a}\right)^{\frac{i\omega}{a}}\Gamma\left(-\frac{i\omega}{a}\right) \nn\\
&-\frac{i}{2\pi k}\sqrt{\frac{\omega}{ k}}e^{-i\omega L} e^{-aL}e^{i\,k/a\,e^{aL}} 
\end{align} 
\\
We can then also evaluate the frequency dependent temperature from $|\beta_{\omega k}^{loc}|^2$ given by, \begin{equation}
T(\omega) = T_R + \frac{e^{-aL}}{k\sqrt{\omega}}\left[T_R(e^{\omega/T_R}-1)\right]^{3/2}\cos\theta 
\end{equation}
\\
where $T_R=a/2\pi$ is the usual Rindler temperature.

\section{Summary}
\label{summary}

\section*{Acknowledgements}
Research of the author is supported by the Core Research Grant CRG/2021/003053 from the Science and Engineering Research Board, India.

\appendix{}

\section{Detailed calculation of \eq{localbeta}}
\label{calcofbeta}

We are required to evaluate the following expression: 
\begin{equation*}
\beta_{\omega k}^{loc} = \frac{i a}{2\pi\sqrt{\omega k}}\lim_{\mu \to 1}\pp_\mu \int_{-\infty}^L \dd v\, e^{-i\omega v}e^{-i (k\mu/a)\, e^{a v}}
\end{equation*}
\\
For this, we substitute $\xi = \frac{k}{a}\mu\,e^{av}$ and with $\tilde{L} = \mu\,k/a\,e^{aL}$ then we have,
\begin{align*}
\beta_{\omega k}^{loc} &= \frac{i a}{2\pi\sqrt{\omega k}}\lim_{\mu \to 1}\pp_\mu \left[\int_0^{\tilde{L}}\frac{\dd \xi}{a}\, e^{-i\xi}\,\xi^{-i\omega/a-1}\,\left(\frac{a}{k\mu}\right)^{-\frac{i\omega}{a}}\right]\nonumber\\
&= \frac{i}{2\pi\sqrt{\omega k}}\left(\frac{k}{a}\right)^{i\omega/a}\lim_{\mu \to 1}\pp_\mu \left[\mu^{i\omega/a}\int_0^{\tilde{L}}\dd \xi\, e^{-i\xi}\,\xi^{-i\omega/a-1}\right]\nonumber\\
&= \frac{i}{2\pi\sqrt{\omega k}}\left(\frac{k}{a}\right)^{\frac{i\omega}{a}}\left[\left(\frac{i\omega}{a}\right)\int_0^{\frac{k}{a}\,e^{aL}}\dd\xi\,e^{-i\xi}\,\xi^{-i\omega/a-1}\right.\nn\\ 
&\left.+ \left(\frac{k}{a}\right)^{-\frac{i\omega}{a}}e^{-i\omega L}e^{-i k/a\,e^{aL}}\right]\nonumber\\
&= \frac{i}{2\pi\sqrt{\omega k}}\left(\frac{k}{a}\right)^{\frac{i\omega}{a}}\left[\left(\frac{i\omega}{a}\right) I + \left(\frac{k}{a}\right)^{-\frac{i\omega}{a}}e^{-i\omega L}e^{-i k/a\,e^{aL}}\right]
\end{align*}
Where the integral $I$ is to be calculated as a contour integral in the complex-$\xi$ plane. This can be done as
\begin{align*}
I &= \int_0^{\frac{k}{a}\,e^{aL}}\dd\xi\,e^{-i\xi}\,\xi^{-i\omega/a-1}\nonumber\\
  &= e^{-\pi\omega/2a}\int_0^{\frac{k}{a}\,e^{aL}}\dd\xi\,e^{-\xi}\,\,\xi^{-i\omega/a -1} - \int_\mathcal{C}\dd \xi\, e^{-i\xi}\,\xi^{-i\omega/a-1}\nonumber\\
  &=  e^{-\pi\omega/2a}\gamma\left(-\frac{i\omega}{a},\frac{k}{a}e^{aL}\right) \nn\\
  &+ i \left(\frac{k}{a}\,e^{aL}\right)^{-i\omega/a} \int_0^{\frac{\pi}{2}}\dd \theta\,e^{-i\,k/a\,e^{aL}\,e^{-i\theta}}\, e^{-\theta\,\omega/a}\nonumber\\
  &= e^{-\pi\omega/2a}\gamma\left(-\frac{i\omega}{a},\frac{k}{a}e^{aL}\right) + i\left(\frac{k}{a}\right)^{-i\omega/a}e^{-i\omega L}I_c
\end{align*}
where $I_c$ is the integral over the contour part. We look at it for large $aL$ limit which is what we are really after. Hence, $\beta_{\omega k}^{loc}$ is given by,
\begin{align*}
\beta_{\omega k}^{loc} =& -\frac{e^{-\pi\omega/2a}}{2\pi a}\sqrt{\frac{\omega}{k}}\left(\frac{k}{a}\right)^{\frac{i\omega}{a}}\gamma\left(-\frac{i\omega}{a},\frac{k}{a}e^{aL}\right)\nn\\
& - \frac{i}{2\pi a}\sqrt{\frac{\omega}{ k}}e^{-i\omega L}I_c + \frac{i}{2\pi\sqrt{\omega k}}e^{-i\omega L}e^{-ik/a\,e^{aL}}.
\end{align*}

\end{document}